\documentstyle[11pt]{article}
\newcommand{\beq}{\begin{equation}}
\newcommand{\eeq}{\end{equation}}
\newcommand{\beqa}{\begin{eqnarray}}
\newcommand{\eeqa}{\end{eqnarray}}

\title{The Zero Temperature Chiral Phase Transition
in $SU(N)$ Gauge Theories}
\author{Thomas Appelquist\\
Department of Physics, Yale University, New Haven, CT 06511
\\ \\
John Terning\\
Department of Physics, Boston University\\
590 Commonwealth Ave., Boston, MA  02215
\\  \\
L.C.R. Wijewardhana \\
Department of Physics, University of Cincinnati, Cincinnati,
OH 45221}
\date{May 28, 1996}

\begin{document}
\setlength{\baselineskip}{24pt}
\maketitle
\begin{picture}(0,0)(0,0)
\put(295,370){YCTP-P2-96}
\put(295,360){BUHEP-96-3}
\put(295,350){UCTP-002-96}
\end{picture}
\vspace{-36pt}

\begin{abstract}
We investigate the zero temperature chiral phase transition in an $SU(N)$
gauge theory as the number of fermions $N_f$ is varied.  We argue that there
exists a critical number of fermions $N_f^c$, above which there is no chiral
symmetry breaking or confinement, and below which both chiral symmetry
breaking and confinement set in. We estimate $N_f^c$ and discuss the
nature of the phase transition.
\end{abstract}

   An $SU(N)$ gauge theory, even at zero temperature,  can exist in different
phases depending on the number of massless fermions $N_f$ in the theory.
The phases are
defined by whether or not chiral symmetry breaking takes
place. For QCD with two or three light quarks, chiral
symmetry breaking and confinement occur at roughly the same scale. By contrast,
in any $SU(N)$
gauge theory, asymptotic freedom (and hence chiral symmetry breaking and
confinement) is lost if the number of fermions is larger than a certain value
($= 11N/2$ for fermions in the fundamental representation).

 If the number of fermions $N_f$ is reduced to just below $11N/2$, an
infrared fixed point will appear, determined by the first two terms in the beta
function. By taking the large $N$ limit
or by continuing to non-integer values of $N_f$ \cite{Banks}, the value of the
coupling at the fixed point can be made
arbitrarily small, making a perturbative analysis reliable.  Such a theory with
a perturbative
fixed point is a massless conformal theory. There is no chiral symmetry
breaking and no confinement.

As $N_f$ is reduced further, chiral symmetry breaking and
confinement will set in.
There have been lattice Monte Carlo studies of the $N_f$ dependence of
chiral symmetry breaking \cite{lattice}.  For example, Kogut and Sinclair
\cite{lattice}
found that for
$N=3$ and $N_f=12$ there is no chiral symmetry breaking, while Brown et. al.
\cite{lattice} have found chiral symmetry breaking for $N=3$ and $N_f = 8$.
In this paper we will estimate the critical value $N_f^c$ at which this
transition occurs. We
then investigate the properties of the phase transition for $N_f \approx
N_f^c$.

  Our discussion will parallel an analysis of the chiral phase transition in
QED3 and QCD3
\cite{QED3,QED3PT}.
In a large $N_f$ expansion it was
found that an appropriate effective coupling has an infrared fixed point with
strength proportional to $1/N_f$, and that as $N_f$ is
lowered, the value of the fixed point exceeds the critical value necessary to
produce
spontaneous chiral symmetry breaking. It was argued that this critical value is
large enough to make the $1/N_f$ expansion reliable.

An  $N_f$ dependence similar to the one we describe here has been found in $N
= 1$ supersymmetric QCD
\cite{Seiberg}.  This theory is not asymptotically free for large enough
$N_f$, and has an infrared, conformal  fixed point  for a range of $N_f $ below
a
certain value.

  The Lagrangian of an $SU(N)$ gauge theory is:
\beq
 {\cal L} = \bar{\psi}(i\not\!\partial + g(\mu)\not\!\!A^a T^a)\psi
+
{1 \over 4} F^a_{\mu\nu}F^{a\mu\nu}
\label{L}
\eeq
where $\psi$ is a set of $N_f$ 4-component spinors, the $T^a$ are
the generators of $SU(N)$, and $g(\mu)$
is the gauge coupling renormalized at some scale $\mu$.
The renormalization group (RG) equation for the running coupling  is:
\beq
\mu{{\partial}\over{\partial \mu}} \alpha(\mu) = \beta(\alpha)
\equiv -b\, \alpha^2(\mu) -c\, \alpha^3(\mu)-d\, \alpha^4(\mu) - ...~,
\label{beta}
\eeq
where $\alpha(\mu) = g^2(\mu)/4 \pi$.  With the $N_f$ fermions in the
fundamental representation, the first two coefficients are given by
\beq
b = {{1}\over{6 \pi}} \left( 11 N - 2 N_f\right)
\label{b}
\eeq
\beq
c = {{1}\over {24  \pi^2}} \left(34 N^2 - 10  N N_f - 3{{N^2 -
1}\over{N}} N_f\right)~.
\label{c}
\eeq
The theory is asymptotically free if $b > 0$ ($N_f < {{11}\over{2}}N$).
At two loops, the theory has
an infrared stable, non-trivial fixed point if $b > 0$ and $c < 0$.  In this
case the fixed point
is at
\beq
\alpha_* = - \,{{ b}\over {c}}~.
\eeq

Recall that the coefficients $b$ and $c$ are scheme-independent \cite{Gross},
while the
higher-order coefficients are scheme-dependent.  In fact one can always choose
a renormalization scheme such that all the higher order coefficients are zero,
i.e.
they  can be removed by a redefinition of
the coupling (change of
renormalization scheme) $g^\prime =  g + G_{1}g^3 + G_{2} g^5+ ...$ .
Thus
if  a  zero, $\alpha_*$, of the $\beta$ function exists at two loops, it exists
to any order in perturbation
theory \cite{Gross}. Of course if the value of  $\alpha_*$  is large enough,
there could be important higher order
corrections to the Green's functions of physical interest. Indeed, their
perturbation expansion might not converge at all.
In addition, non-perturbative effects, such as
spontaneous chiral symmetry breaking, could eliminate even the existence of the
fixed point.
If the quarks
develop a dynamical mass, for example, then below this scale only
gluons
will contribute to the $\beta$ function, and the perturbative fixed point
turns out to be only an approximate description, relevant above the chiral
symmetry breaking scale.

  For $N_f$ sufficiently close to $11N/2$, the value of the coupling at the
infrared fixed point can be made arbitrarily small.  The RG equation for the
running coupling can be written as
\beq
b\log\left({{q}\over{\mu}}\right) = {{1}\over{\alpha}} -
{{1}\over{\alpha(\mu)}}- {{1}\over{\alpha_*}}\log\left({{\alpha
\left(\alpha(\mu) -
\alpha_*\right)}\over{\alpha(\mu)\left(\alpha-\alpha_*\right)}}\right)~,
\eeq
where $\alpha = \alpha(q)$.  For $\alpha$, $\alpha(\mu)<\alpha_*$ we can
introduce a scale defined by
\beq
\Lambda = \mu \exp\left[{{-1}\over{b \,\alpha_*}}
\log\left({{\alpha_*-\alpha(\mu)}
\over{\alpha(\mu)}}\right)-{{1}\over{b \alpha(\mu)}}\right]~,
\label{Lambda}
\eeq
so that
\beq
{{1}\over{\alpha}} = b \log\left({{q}\over{\Lambda}}\right) +
{{1}\over{\alpha_*}}
\log\left({{\alpha}\over{\alpha_*-\alpha}}\right).
\eeq
Then for $q \gg \Lambda$ the running coupling displays the usual
perturbative behavior:
\beq
\alpha \approx {{1}\over{b \log\left({{q}\over{\Lambda}}\right)}}~,
\label{highalpha}
\eeq
while for $q \ll \Lambda$ it approaches the fixed point $\alpha_*$:
\beq
\alpha \approx {{\alpha_*}\over{1+ {{1}\over{e}}
\left({{q}\over{\Lambda}}\right)^{b \alpha_*}}}~.
\label{lowalpha}
\eeq

 As $N_f$ is decreased, the infrared fixed point $\alpha_*$ increases. We will
suggest here that the breakdown of perturbation theory, described above, first
happens due to the spontaneous breaking of chiral symmetry, and that the phase
transition can be described by
an RG improved ladder approximation of the CJT \cite{CJT} effective potential.
It is well known \cite{Peskin} that in vector-like gauge theories
the two-loop effective potential
expressed
as a functional of the quark self-energy becomes unstable to chiral symmetry
breaking
when the gauge coupling exceeds a critical value\footnote{A more general
definition
of the critical coupling is that the anomalous dimension of
$\overline{\psi}\psi$ becomes 1 \cite{rainbow}.} :
\beq
\alpha_c \equiv {{ \pi }\over{3 \, C_2(R)}}= {{2 \pi \,
N}\over{3\left(N^2-1\right)}}~,
\label{alphacrit}
\eeq
where $C_2(R)$ is the quadratic Casimir of the the representation $R$.
Thus we would expect that when $N_f$ is decreased below the value $N_f^c$ at
which $\alpha_{*} = \alpha_c$, the theory
undergoes a  transition to a phase where chiral symmetry is
spontaneously broken. The critical value $N_f^c$ is given by

\beq
N_f^c = N \left({{100N^2 -66}\over{25 N^2 -15}}\right).
\label{Ncrit}
\eeq
For large $N$, $N_f^c$ approaches $4N$, while for $N=3$, $N_f^c$ is just below
12.
Note that this is consistent with lattice QCD results \cite{lattice},
which suggest that $8 < N_f^c \le 12$.

Is this simple analysis reliable? After all, it could be that when $\alpha_*$
is as large as $\alpha_c$ the
perturbative expansion for the CJT potential has broken down. To address this
question we provide a crude
estimate of the higher order corrections to the CJT potential. An explicit
computation of the next-to-leading term (or equivalently the next-to-leading
term in the gap
equation) \cite{ALM} for $\alpha_*  \approx \alpha_c$, produces an additional
factor of approximately
$\epsilon = {{\alpha_{c}N}\over{4 \pi}}$.
This is the factor remaining after the appropriate renormalizations are
absorbed into the definition of the
coupling constant.  From equation (\ref{alphacrit}) we see that
\beq
\epsilon = {{1}\over{6 \left(1- {{1}\over{N^2}}\right)}}~.
\eeq
For QCD, $\epsilon \approx 0.19$. If higher orders in the computation
produce approximately this factor, the perturbative
expansion of the CJT potential may be reliable\footnote{It is worth noting that
in condensed matter physics one can often (though not always) obtain useful
information from the Wilson-Fisher expansion in a parameter
that is set to one at the end
of the calculation.}. The same may be true of the various Green's functions
encountered in the skeleton expansion of the CJT potential.

We next explore the nature of the chiral phase transition at $N_{f} = N_f^c$
and its relation to confinement. It is useful to consider first the behavior in
the broken phase  $ N_f  < N_f^c$  ( $\alpha_{*} > \alpha_c$).
Here each quark develops a dynamical mass $\Sigma(p)$. For $N_f  \rightarrow
N_f^c$ from below ($\alpha_{*} \rightarrow \alpha_c$ from above), $\Sigma(p)$
can be determined by solving a linearized Schwinger-Dyson gap equation in
ladder approximation. For momenta small compared to $\Lambda$,  the effective
coupling strength is $\alpha_*$, while for momenta above $\Lambda$ it falls
according to equation (\ref{highalpha}). The resulting solution for $\Sigma(0)$
is \cite{SD}

\beq
\Sigma(0) \approx \Lambda
\exp\left({{- \pi}\over{\sqrt{{{\alpha_*}\over{\alpha_c}} -1}}}\right) ~.
\label{critical}
\eeq
The  behavior of $\Sigma(p)$ as a function of $p$ will be discussed shortly.

Once the dynamical mass $\Sigma(p)$ is formed, the fermions decouple below this
scale, leaving the pure gauge theory behind. One might worry that this would
invalidate the above gap equation analysis since it relies on the fixed point
which only exists when the fermions  contribute to the $\beta$ function.  This
is not a problem, however, since it can be shown that when
$\Sigma(0) \ll \Lambda$ the dominant momentum
range in the gap equation, leading to the exponential behavior of equation
(\ref{critical}), is $\Sigma(0) < p < \Lambda$. In this range, the fermions are
effectively massless and the coupling does appear to be approaching an infrared
fixed point.  Note that the condition $\Sigma(0) \ll \Lambda$ is indeed
satisfied for
$N_f$ sufficiently close to $N_f^c$.

Below the scale $\Sigma(0)$ the quarks can be integrated out; thus the
effective $\beta$ function has no fixed point and the gluons are confined. The
confinement scale can be estimated by noting that at the quark decoupling scale
$\Sigma(0)$, the effective coupling constant is of order $\alpha_c$. A simple
estimate using equations (\ref{beta})-(\ref{c}) then reveals that
the confinement scale is roughly the same order of magnitude
as the chiral symmetry breaking scale.  When $N_f $ is reduced sufficiently
below $N_f^c$ so that $\alpha_*$ is not close to $\alpha_c$,  both $\Sigma(0)$
and the confinement scale become of order $\Lambda$. The linear approximation
to the gap equation will then no longer be valid, and it will probably no
longer  be the
case that higher order contributions to the effective potential can be argued
to be small.

It is interesting to compare the behavior of the broken phase
for $N_f$ near $N_f^c$ to the walking technicolor gauge theories
discussed recently in the literature \cite{walking}. We have argued here that
for $N_f $ just below $N_f^c$, the dynamical breaking is governed by a
linearized ladder gap equation with a coupling $\alpha_*$ just above
$\alpha_c$. As the momentum $p$ increases,
$\alpha(p)$ stays near $\alpha_*$ (it ``walks") until $p$ becomes of
order $\Lambda$, and only falls above this scale.
It can then be seen \cite{TABrazil} that the dynamical mass $\Sigma(p)$ falls
like $1/p$ (i.e. the anomalous dimension of
$\overline{\psi}\psi$ is $\approx1$) for $\Sigma(0) < p < \Lambda $ and only
begins to
fall more rapidly ( like $1/p^2$ ) at larger momenta. This is precisely the
walking behavior employed in technicolor theories and referred to there as high
momentum enhancement. In that case, however, there was no IR fixed point to
keep the $\beta$ function near zero and slow the running of the coupling. It
was noted instead that the same effect would emerge if the $\beta$ function was
small at each order by virtue of partial cancelations between fermions and
bosons.

  From the smooth behavior of the order parameter $\Sigma(0)$ (equation
(\ref{critical})), it would naively appear that the  chiral phase transition
at $N_f = N_f^c$ ($\alpha_{*} = \alpha_{c}$) is second order. In this paper
we will use the phrase ``second order"  to refer exclusively
to a phase transition where the correlation
length diverges as the critical point is approached from either side.
In other words,
there is a light excitation coupling to the order
parameter that becomes massless at the critical point. In the broken phase,
this mode would be present along with the massless
Goldstone modes. In the symmetric phase, all these modes would form
a light, degenerate multiplet,
becoming massless at the critical point \cite{NJL}.

  We examine the correlation length by working in the symmetric phase and
searching for poles  in the (flavor and color-singlet) quark-antiquark
scattering amplitude, computed in the same (RG improved, ladder)
approximation leading to equation (\ref{critical}).
The analysis is similar to that carried out for QED3 \cite{QED3PT}.
If the transition
is  second-order, then at least one pole should move to zero
momentum
as we approach
the critical point (i.e. the correlation length should diverge).
We take the incoming (Euclidean) momentum of the initial
quark and
antiquark to be $q/2$, but
keep a non-zero momentum transfer by assigning outgoing momenta
$q/2 \pm p$ for the final quark
and antiquark. Any light scalar resonances should make their presence
known by producing a pole in the scattering amplitude (when continued to
Minkowski
$q^2$).

If the Dirac indices of the initial quark and antiquark are $\lambda$ and
$\rho$, and  those of the final state quark and antiquark are $\sigma$
and $\tau$, then the scattering
amplitude can be written (for small $q$) as
$T_{\lambda \rho \sigma \tau}(p,q) = \delta_{\lambda \rho} \delta_{\sigma
\tau}\,T(p,q)/p^2 + . . .$,
where the ... indicates pseudoscalar, vector, axial-vector, and tensor
components, and we have factored out $1/p^2$ to make $T(p,q)$
dimensionless.   We contract
Dirac indices so that we obtain the Bethe-Salpeter equation for the the scalar
s-channel scattering amplitude $T(p,q)$, containing only t-channel
gluon exchanges.  If $p^2 \gg q^2$, then $q^2$ will simply act as an
infrared cutoff in the loop integrations.
The Bethe-Salpeter equation  in the scalar channel for $p \ll \Lambda$
is:
\begin{eqnarray}
T(p,q)  \approx  {{\alpha_*}\over{\alpha_c}} \pi^2+
{{\alpha_*} \over{4\alpha_c}}\,\int_{q^2}^{p^2} dk^2 \,T(k,q) \,{{
1}\over{k^2}} +
{{\alpha_*} \over{4\alpha_c}}\,\int_{p^2}^{\Lambda^2} dk^2 \,
T(k,q) \,{{ p^2}\over{k^4}},
\label{Tstart}
\end{eqnarray}
where $\Lambda$ is the scale introduced in equation (\ref{Lambda}).
 (Note that contributions from the
integration region $k^2 > \Lambda^2$ are suppressed by a factor
$p^2/\Lambda^2$,
and a falling $\alpha(k)$.)
The first term in equation (\ref{Tstart}) is simply one gluon
exchange. We have used Landau gauge ($\xi=1$) where the quark
wavefunction renormalization vanishes to lowest order.
Because of the existence of the fixed point, it is a
good approximation  to have replaced $\alpha(p)$~and
$\alpha(p-k)$~by~$\alpha_*$~at momentum scales below $\Lambda$.

For momenta $p^2>q^2$, equation~(\ref{Tstart}) can be converted to a
differential equation
with appropriate boundary conditions.
The solutions have the form
\begin{equation}
T(p,q) = A(q)
\left({p^2 \over \Lambda^2}\right)^{{1 \over 2} + {1 \over 2} \eta}
 +B(q)
\left({p^2 \over \Lambda^2}\right)^{{1 \over 2} - {1 \over 2} \eta} ~,
\label{sol}
\end{equation}
where $\eta= \sqrt{1-\alpha_*/\alpha_c}$.
The coefficients $A$ and $B$ can be
determined by substituting this solution back into equation~(\ref{Tstart}).
This
gives:
\begin{equation}
A ={{-2 \pi^2\left(1-\eta\right)^2}\over{\left(1+\eta\right)}}
{{\left({{q^2}\over {\Lambda^2}}\right)^{-{1\over 2}+{1\over 2}\eta} }
\over { 1-
\left({{1-\eta}\over{1+\eta}}\right)^2
\left({{q^2}\over {\Lambda^2}}
\right)^\eta }   }
{}~,
\label{A}
\end{equation}
and
\begin{equation}
B =
{{2 \pi^2\left(1-\eta\right)
\left({{q^2}\over {\Lambda^2}}\right)^{-{1\over 2}+{1\over 2}\eta} }
\over { 1-
\left({{1-\eta}\over{1+\eta}}\right)^2
\left({{q^2}\over {\Lambda^2}}
\right)^\eta }   }~.
\label{B}
\end{equation}
Note that there is an infrared divergence in the limit $q^2 \rightarrow 0$ in
both
equations (\ref{A}) and (\ref{B}).  That this is an infrared divergence rather
than a pole corresponding to a bound state can be seen from the fact that
the divergence exists for arbitrarily weak coupling ($\alpha_* \rightarrow 0$).
In fact, it can already be seen at order $\alpha_*^2$ in the
one-loop (two
gluon exchange) diagram.
As required by the KLN theorem \cite{KLN}, this infrared divergence will be
cancelled in a physical
scattering process by the emission of soft quanta.

If we denote the location of the poles of the functions $A$ and $B$ in the
complex $q^2$ plane by $q_0^2$, we have
\begin{equation}
|q_0^2| = \Lambda^2
\left({1+\eta}\over
{1-\eta}\right)^{2 \over \eta}~.
\label{q0}
\end{equation}
We see that there is no pole that approaches the origin $q_0^2 = 0$
as $\alpha_* \rightarrow \alpha_c$.
Thus
the correlation length does not diverge, and  the transition is not
second order (It is not  conventionally first order  either since the
order parameter vanishes continuously at the critical point.). Note
that the behavior of
the zero temperature chiral phase transition is different from the
finite temperature case due to the presence of long-range gauge forces.  At
finite temperatures, gluons are
screened, and thus there are only short-range forces present and only
conventional first or second order transitions are possible.

To conclude, we have argued that as the number of quark flavors, $N_f$, is
reduced, QCD-like
theories in four dimensions
undergo a chiral phase transition at a critical value $N_f^c$ (equation
(\ref{Ncrit})). For
$N_f<N_f^c$, chiral
symmetries are spontaneously broken, while they are unbroken for
$N_f>N_f^c$. We have explored the nature of the chiral phase transition,
arguing that it can be described using the QCD gap equation in ladder
approximation (equivalently the two-loop approximation to the CJT potential).
We have
also argued that even though the order parameter vanishes at the critical
point, the correlation length does not diverge
(i.e. the phase transition is not  second order). The critical
behavior described here is similar to that found in QED3 and QCD3
\cite{QED3,QED3PT}. We have, of course, not proven that higher order
corrections to our computation are small. Further study of this question as
well as lattice Monte Carlo studies of the zero temperature phase transition
would help to confirm or disprove our conclusions.

\noindent \medskip\centerline{\bf Acknowledgements}
\vskip 0.15 truein
We thank  S. Chivukula, A. Cohen, P. Damgaard, S. Hsu, M. Luty, and R. Sundrum
for
helpful discussions.
This work was supported in part by  the Department of Energy under
contracts \#DE-FG02-92ER40704 and \#DE-FG02-91ER40676,  \#DE-FG-02-84ER40153.
\vskip 0.15 truein


\end{document}